\documentstyle [12pt] {article}

\newcommand{\gtwid}{\mathrel{\raise.3ex\hbox{$>$\kern-.75em\lower1ex
\hbox{$\sim$}}}}
\newcommand{\ltwid}{\mathrel{\raise.3ex\hbox{$<$\kern-.75em\lower1ex
\hbox{$\sim$}}}}
\newcommand{\beq}{\begin{equation}}
\newcommand{\eeq}{\end{equation}}
\newcommand{\beqs}{\begin{eqnarray}}
\newcommand{\eeqs}{\end{eqnarray}}

\catcode`@=11
\@addtoreset{equation}{section}
\@addtoreset{equation}{subsection}
\def\theequation{\ifnum\value{section}=0 \arabic{equation}\ignorespaces
\else \ifnum\value{section}=-1 A.\arabic{equation}\ignorespaces
\else \ifnum\value{subsection}=0 \thesection.\arabic{equation}\ignorespaces
\else \thesection.\arabic{subsection}.\arabic{equation}\ignorespaces
                           \fi
                      \fi
                 \fi}
\catcode`@=12

\begin{document}

\def\thefootnote{\fnsymbol{footnote}}
\baselineskip 6.0mm

\begin{flushright}
\begin{tabular}{l}
ITP-SB-95-52    \\
December, 1995 \\
\end{tabular}
\end{flushright}

\vspace{8mm}

\begin{center}

{\Large \bf Implications of Anomaly Constraints in the}

\vspace{4mm}

{\Large \bf $N_c$--Extended Standard Model }

\vspace{12mm}

\setcounter{footnote}{0}
Robert Shrock\footnote{email: shrock@insti.physics.sunysb.edu}

\vspace{6mm}
Institute for Theoretical Physics  \\
State University of New York       \\
Stony Brook, N. Y. 11794-3840  \\

\vspace{16mm}

{\bf Abstract}
\end{center}
    We discuss some implications of anomaly cancellation in the
standard model with (i) the color group extended to $SU(N_c)$, and (ii) the
leptonic sector extended to allow right-handed components for neutrinos.

\vspace{16mm}

\pagestyle{empty}
\newpage

\pagestyle{plain}
\pagenumbering{arabic}
\renewcommand{\thefootnote}{\arabic{footnote}}
\setcounter{footnote}{0}

\section{Introduction}

    While the standard model \cite{sm}
is quite successful in explaining known data,
there are many questions which it leaves unanswered.  One of the most basic is
why the gauge group is $G_{SM} = SU(3) \times SU(2) \times U(1)_Y$.  There have
been several appealing ideas which might answer, or help to answer, this
question, such as
grand unification and the more ambitious efforts to derive the standard model
 from a fundamental theory of all known interactions, including
gravity.\cite{kk} A somewhat complementary approach is to consider how the
standard model gauge group might be viewed as one of a sequence of gauge
groups (which, in general, are still products of factor groups).
In particular, it has proved quite useful to consider the number of
colors as a parameter, and study the $N_c \to \infty$ limit of the quantum
chromodynamics (QCD) sector of the theory \cite{thooft}-\cite{wittennc},
since this enables one to carry out analytic nonperturbative calculations
(and, indeed, to obtain a soluble model in $d=2$ spacetime
dimensions).\cite{other}  Many of these discussions
naturally concentrated on using the $1/N_c$
expansion to elucidate the properties of hadrons.  When one includes
electroweak interactions, however, one is led to address some additional
questions.  One of these concerns anomalies.

   The freedom from anomalies is a necessary property of an acceptable
quantum field theory.  In $d=4$ dimensions, there are three types of possible
anomalies in quantum field theories, including
(i) triangle anomalies in gauged currents \cite{abj,gj} which,
if present, would spoil current conservation and hence renormalizability; and
(ii) the global SU(2) anomaly resulting from the nontrivial homotopy group
$\pi_4(SU(2))=Z_2$ \cite{witten} which, if present, would render the path
integral ill-defined.  Furthermore, (iii) if one includes
gravitational effects on a semi-classical, even if not fully quantum level,
one is motivated to require the absence of mixed gauge-gravitational
anomalies \cite{gaugegrav} resulting from triangle diagrams involving two
energy momentum tensor (graviton) vertices and a $U(1)_Y$ gauge vertex, since
this anomaly, if present, would also spoil conservation of the hypercharge
current as well as precluding the construction of a generally covariant
theory.  As is well known, in the
standard model (SM), all of these anomalies vanish
\cite{sm,bim,gj}, and for anomalies of type (i) and (ii), this
vanishing occurs in a manner which intimately connects  the
quark and lepton sectors.  Furthermore, in the standard model (with no
right-handed neutrinos) the cancellation of the anomalies of type (i)
implies the quantization of the fermion electric charges \cite{marsh}; this
also holds in an extension of the standard model where right-handed
neutrinos are included but
are assumed to have zero hypercharge \cite{marsh}.  Note that the
gauge-gravitational anomaly vanishes separately for quark and lepton sectors.

   The issue of anomalies in the $N_c$--extended standard model has recently
been addressed explicitly by Chow and Yan \cite{cy}.
These authors note that the anomaly cancellation conditions can
be satisfied for arbitrary (odd) $N_c$, and the solution leads to unique,
quantized, values of the electric charges of the up-type and down-type
quarks, $q_u$ and $q_d$.
The present author had carried out a
similar analysis for a different type
of generalization of the standard model, namely, one in which the color group
is extended to $SU(N_c)$ and the leptonic sector is extended to include
right-handed neutrino fields.

    In this paper, we shall discuss the results of this analysis.  These
results present an interesting contrast to those in
the $N_c$--extended standard model (with no right-handed neutrinos).
Both types of generalizations of the SM (excluding or including
right-handed neutrinos) are of interest. The generalization
without any right-handed neutrinos may provide a more economical way of
getting small neutrino masses (via dimension-5 operators \cite{dim5}), while
the generalization with right-handed neutrinos
is motivated in part by the fact that these make possible Dirac and
right-handed Majorana mass terms for neutrinos at the renormalizable,
dimension-4 level, which naturally yield small observable neutrino
masses via the seesaw mechanism \cite{seesaw}
\cite{nutc}, given that the natural scale for the mass coefficients
of the right-handed Majorana neutrino bilinears is much larger than the
electroweak symmetry breaking (EWSB) scale.  In the usual extension of
the standard model, the right-handed neutrino fields are electroweak
singlets, a property which is crucial for the existence of the right-handed
Majorana mass term. However, when one considers the $N_c$--extended standard
model with right-handed neutrino fields from the perspective of determining
the constraints on the fermion hypercharges $Y_f$ which
follow from the requirement of cancellation of anomalies, the hypercharge of
the right-handed neutrinos (like the hypercharges of the other fields)
naturally becomes a variable, not necessarily equal to zero \cite{add}.
If the hypercharge, and hence electric charge, of the right-handed
neutrinos is nonzero, then the nature of the theory changes in a
fundamental way.  Indeed, the term ``neutrino'' becomes a misnomer; we shall
retain it here only to avoid proliferation of terms (it is no worse than the
accepted term ``heavy lepton'').  Clearly, if
$Y_{\nu_R} \ne 0$, then the right-handed Majorana bilinear
$\nu_{iR}^T C \nu_{jR}$ is forbidden by gauge invariance
(where $i,j$ denote generation indices, and $C$ denotes the Dirac
charge conjugation matrix).  Given that Dirac mass terms for the neutrinos
would be present in this type of theory, it would be natural for all of the
fermions of a given generation to have comparable masses \cite{lep}.  This
class of models is of interest from an abstract field-theoretic viewpoint,
because it serves as a theoretical laboratory in which to investigate the
properties that follow from anomaly cancellation in a chiral gauge theory
constituting a generalization of the standard model with $N_c$ colors,
constructed such that all left-handed Weyl components have right-handed
components of the same electric charge.

In most of our discussion, we shall not need to make any
explicit assumption concerning the still-unknown origin of electroweak
symmetry breaking.  At appropriate points, we shall comment on how various
formulas would apply in the $N_c$-extended minimal supersymmetric standard
model (MSSM) as well as the $N_c$-extended standard model itself (in both
cases, including right-handed components for all matter fermions).  As regards
anomalies in the context of the MSSM, recall that in addition to the usual
Higgs $H_d$, one must introduce another, $H_u$, with opposite hypercharge, both
in order to be able to give the up-type quarks masses while maintaining a
holomorphic superpotential, and in order to avoid anomalies in gauged
currents which would be caused by the higgsino $\tilde H_d$ if it were not
accompanied by a $\tilde H_u$.  (The addition of a single $\tilde H_d$ to the
(even) number of matter fermion SU(2) doublets would also cause a global
SU(2) anomaly.)  All of this works in the same way regardless of the charges of
the matter fermions, provided that the latter satisfy the anomaly cancellation
condition by themselves.  Moreover, as
regards the neutralino sector, if $q_\nu=0$, then electric charge conservation
by itself would allow mixing of neutrinos and neutralinos (the neutral
higgsinos and superpartners of the gauge fields $\tilde A^0$ and $\tilde B$) if
and only if $q_\nu=0$.  However, the $R$ parity commonly invoked in the MSSM
to prevent disastrously rapid proton decay also prevents mixing among the
neutralinos and neutrinos even in the conventional case where $q_\nu=0$, so
there would be no change concerning this mixing even if $q_\nu \ne 0$.
Finally,
considering alternative ideas for electroweak symmetry breaking, one could
envision embedding the $G_{SM}'$ theory in a
larger one in which this symmetry breaking is dynamical.

\section{Anomaly Constrants and Their Implications}
\subsection{General}

     Consider, then, the generalization
\beq
G_{SM} \to  G_{SM}' = SU(N_c) \times SU(2) \times U(1)_Y
\label{group}
\eeq
with the fermion fields consisting of the usual $N_{gen.}=3$
generations, each containing the following representations of $G_{SM}'$:
\beq
Q_{iL} = \left (\begin{array}{c}
                  u_i \\
                  d_i \end{array} \right )_L \ : \ (N_c,2,Y_{Q_L})
\label{ql}
\eeq
\beq
u_{iR} \ : \ (N_c,1,Y_{u_R})
\label{ur}
\eeq
\beq
d_{iR} \ : \ (N_c,1,Y_{d_R})
\label{dr}
\eeq
\beq
{\cal L}_{iL} = \left (\begin{array}{c}
                  \nu_i \\
                   e_i \end{array} \right )_L \ : \ (1,2,Y_{{\cal L}_L})
\label{ll}
\eeq
\beq
\nu_{jR} \ : \ (1,1,Y_{\nu_R})
\label{nur}
\eeq
\beq
e_{iR} \ : \ (1,1,Y_{e_R})
\label{er}
\eeq
where the index $i$ denotes generation, $i=1,..N_{gen.}=3$, with
$u_1 = u$, $u_2=c$, $u_3=t$, $d_1=d$, $d_2=s$, $d_3=b$, etc. Thus, as usual,
all generations have the same gauge quantum numbers.  (In some
formulas, we shall leave $N_{gen.}$ arbitrary for generality.)
Because the SU(2) representations are the same in $G_{SM}'$ as they were in
$G_{SM}$, the usual relations $Q=T_3+Y/2$, $Y_{Q_L}=q_u+q_d$, $q_u=q_d+1$,
$q_\nu=q_e+1$, and $Y_{f_R}=2q_{f_R}$ continue to hold, independent of the
specific values of the fermion electric charges (where we
have used the vectorial nature of the electric charge coupling,
$q_{f_L}=q_{f_R}=q_f$ for all fermions $f$); and just as in the standard model
itself, these relations imply
\beq
Y_{u_R}=Y_{Q_L}+1 \ , \quad Y_{d_R}=Y_{Q_L}-1
\label{yqcon}
\eeq
and
\beq
Y_{\nu_R}=Y_{{\cal L}_L}+1 \ , \quad Y_{e_R}=Y_{{\cal L}_L}-1
\label{ylcon}
\eeq
Before imposing the anomaly cancellation conditions, there are thus only two
independent electric charges among the fermions; we may take these to be
$q_d$ and $q_e$.  For $N_c=3$ and $q_\nu=0$, one may, {\it a
priori}, have $j=1,...N_s$ electroweak-singlet right-handed
neutrinos $\nu_{jR}$, where $N_s$ need not be equal to $N_{gen.}$.  However,
in the general solution to the anomaly cancellation conditions for
$N_c \neq 3$ (see below) the electric charges of all of the fermions will
differ from their $N_c=3$ values.  In particular, since $q_{\nu}$ will not, in
general, be equal to zero, the number $N_s$ of electroweak-singlet
right-handed neutrinos $\nu_{jR}$ must be equal to the number $N_{gen.}$
of left-handed lepton doublets in order to construct
renormalizable, dimension-4 neutrino mass terms, which in turn is necessary in
this case to avoid massless, charged, unconfined fermions in the theory.  Given
that $N_s=N_{gen.}$, the number $N_{gen.}$ enters in a trivial way as a
prefactor in all of the expression for the anomalies of type (i) and (iii),
i.e. these cancel separately for each generation of fermions.  Accordingly, we
shall often suppress the generational index in the notation henceforth.

 The hypercharge relations (\ref{yqcon}) and (\ref{ylcon})
guarantee that, independent of the specific values of the fermion charges,
one can write $G_{SM}'$-invariant Yukawa couplings
\beq
-{\cal L}_{Yuk} = \sum_{i,j}\biggl [
\Bigl (Y^{(d)}_{ij}\bar Q_{iL} d_{jR} +
       Y^{(\ell)}_{ij}\bar {\cal L}_{iL}e_{jR} \Bigr )H_d
+\Bigl (Y^{(u)}_{ij}\bar Q_{iL}u_{jR} +
        Y^{(\nu)}_{ij}\bar {\cal L}_{iL}\nu_{jR} \Bigr )H_u \biggr ] + h.c.
\label{yuk}
\eeq
where in a context in which one uses a single standard-model Higgs field,
$\phi$, with $I_\phi=1/2$, $Y_{\phi}=1$, then $H_d=\phi$
and $H_u=i\sigma_2 \phi^*$ as usual, and in the minimal
supersymmetric standard model (MSSM), $H_d$ and $H_u$ correspond to the
scalar components of the two oppositely charged Higgs chiral superfields.
(In eq. (\ref{yuk}), no confusion should result between the symbols
${\cal L}_{Yuk}$ for the Lagrangian terms and ${\cal L}_{iL}$ for the
lepton doublets.)  The vacuum expectation value(s) (vevs) of the Higgs fields
then yield fermion mass terms.  We denote these vev's as
$\langle \phi \rangle =
2^{-1/2}v$ in the SM, with $v=2^{-1/4}G_F^{-1/2}$, and $\langle H_{u,d} \rangle
= 2^{-1/2}v_{u,d}$ in the MSSM, with $\tan\beta = v_u/v_d$ and
$v=\sqrt{v_u^2+v_d^2}$. In a scenario without Higgs, in which the electroweak
symmetry breaking is dynamical, the fermion mass terms are envisioned to
arise from four-fermion operators (the origin of which is explained with
further theoretical inputs). In all
three cases, this can be done just as in the respective $N_c=3$ model with
conventional fermion charge assignments.
It is also straightforward to see that in either a
non-supersymmetric model with the single Higgs $\phi$, or the MSSM, or a model
with dynamical electroweak symmetry breaking, the breaking pattern
\beq
G_{SM}' \to SU(N_c) \times U(1)_{em}
\label{symbrk}
\eeq
can be arranged, just as for the $N_c=3$ case with conventional fermion
charges.

\subsection{Anomalies in Gauged Currents}

    We proceed to analyze the constraints from the cancellation of the three
types of anomalies.  Among the triangle anomalies of type (i), the
$SU(N_c)^3$
and $SU(N_c)^2U(1)_Y$ anomalies vanish automatically (as for $N_c=3$)
because of the vectorial nature of the color and electromagnetic couplings.
The condition for the vanishing of the $SU(2)^2 U(1)_Y$ anomaly is
\beq
N_c Y_{Q_L} + Y_{{\cal L}_L}=0
\label{yeq}
\eeq
i.e.,
\beq
N_c(2q_d+1)+(2q_e+1)=0
\label{chargerel}
\eeq
The $U(1)_Y^3$ anomaly vanishes if and only if
\beq
N_c(2Y_{Q_L}^3 - Y_{u_R}^3 - Y_{d_R}^3) +
(2Y_{{\cal L}_L}^3 - Y_{\nu_R}^3 - Y_{e_R}^3) = 0
\label{u1cubed}
\eeq
Expressing this in terms of $q_d$ and $q_e$ yields the same condition as
eq. (\ref{chargerel}).  Solving (\ref{chargerel}) for $q_d$ yields
\beq
q_d = -\frac{1}{2}\biggl ( 1 + \frac{1}{N_c}(2q_e+1) \biggr )
\label{qdsol}
\eeq
and hence
\beq
q_u = \frac{1}{2}\biggl ( 1 - \frac{1}{N_c}(2q_e+1) \biggr )
\label{qusol}
\eeq
or equivalently, taking $q_d$ as the independent variable,
\beq
q_e = -\frac{1}{2}\biggl ( 1 + N_c(2q_d+1) \biggr )
\label{qesol}
\eeq
and thus
\beq
q_\nu = \frac{1}{2}\biggl ((1-N_c(2q_d+1)\biggr )
\label{qnusol}
\eeq

\subsection{Global SU(2) Anomaly}

   The constraint from the global SU(2) anomaly is well-known \cite{witten}:
the number $N_d$ of SU(2) doublets must be even:
\beq
N_d = (1+N_c)N_{gen.} \ \quad is \ \ even
\label{global}
\eeq
For odd $N_{gen.}$, this implies that $N_c$ is odd.  For a nontrivial
color group, this means $N_c=2s+1$, $s\ge 1$.  Note that one gets a
qualitatively different result in the hypothetical case in which $N_{gen.}$ is
even; here, there is no restriction on whether $N_c$ is even or odd.  From a
theoretical point of view, one could perhaps regard it as satisfying that the
physical value $N_{gen.}=3$ is odd and hence is such as to yield a constraint
on $N_c$.\cite{n4}
Of course, a world with even $N_c$ would be very different from our
physical world, since baryons would be bosons.

\subsection{Mixed Gauge-Gravitational Anomalies}

   Finally, the anomalies of type (iii) do not add any further constraint;
the mixed gauge-gravitational anomaly involving $SU(N_c)$ and $SU(2)$ gauge
vertices vanish identically since $Tr(T_a)=0$ where $T_a$ is the generator of a
nonabelian group, and the anomaly involving a $U(1)_Y$ vertex is proportional
to
\beq
N_c(2Y_{Q_L}-Y_{u_R}-Y_{d_R})+(2Y_{{\cal L}_L}-Y_{\nu_R}-Y_{e_R}) = 0
\label{ggu1}
\eeq
where the expression vanishes because of the vectorial nature of the
electromagnetic coupling.  Indeed, the two separate terms in parentheses each
vanish individually: $2Y_{Q_L}-Y_{u_R}-Y_{d_R}=0$ and
$2Y_{{\cal L}_L}-Y_{\nu_R}-Y_{e_R}=0$, so that this anomaly does not connect
quark and lepton sectors, unlike (\ref{yeq}), (\ref{u1cubed}) and the global
SU(2) anomaly.  Hence, the only constraint on the fermion charges is
provided by the condition that the anomalies of type (i) vanish.

\subsection{Discussion}

   Our results show that the SM has a consistent generalization to the gauge
group $G_{SM}'$ in eq. (\ref{group}) with fermion charges given by
(\ref{ql})-(\ref{nur}).  We find the one-parameter family of solutions
given in (\ref{chargerel}) to
the condition of zero anomalies in gauged currents.  Since the values of $q_d$
and $q_e$ for which (\ref{chargerel}) is satisfied are, in general, real, and
are not restricted to the rational numbers,
it follows that in this generalization of
the standard model, the anomaly cancellation conditions do not imply the
quantization of electric charge (and hence, hypercharge).  We note that this is
qualitatively different from the type of generalization studied in
Ref. \cite{cy}, in which one extends $G_{SM} \to G_{SM}'$ but keeps the fermion
content precisely as in the standard model, with no electroweak-singlet
right-handed neutrinos.  In that case, one must keep $q_\nu=0$ in order to
avoid a massless, charged, unconfined fermion,
and hence the lepton charges must be kept at their $N_c=3$ values
while the quark charges are allowed to vary.
Hence, the one-parameter family of solutions (\ref{chargerel}) reduces to a
unique solution
\beq
q_d= q_u-1 = \frac{1}{2}\Bigl ( -1 + \frac{1}{N_c} \Bigr )
\label{qdspecial}
\eeq
and the anomaly cancellation conditions
(specifically, of type (i)) do imply charge quantization, as was noted in
Ref. \cite{cy}.   In passing, we observe that for our type of generalization,
although the generic situation for the solutions of eq. (\ref{chargerel}) is
that $Y_{Q_L}$ and $Y_{{\cal L}_L}$ are real numbers, it is true that this
equation implies that if either is rational, so is the other.

\section{Classification of Solutions for Quark Charges}

   There is another important difference in the properties of the two types of
$N_c$--extended standard model in which one includes or excludes
right-handed neutrinos. In the case where one excludes them,
eq. (\ref{qdspecial}) shows that (given a nontrivial color group)
$q_d$ is always negative, and $q_u$ is always positive, and both decrease
monotonically as functions of $N_c$ (from $(q_u,q_d)=(2/3,-1/3)$ at $N_c=3$
to $(1/2,-1/2)$ in the limit as $N_c \to \infty$).
The situation is qualitatively different in the $N_c$--extended standard model
with right-handed neutrinos; here, there are a number of different
cases (denoted $Cn_q$) describing the up and down quark charges, of which
three are generic and two are borderline.  (We also list a certain special
subcase because of its symmetry.) Regarding
$q_e$ as the independent variable in the solution of eq. (\ref{chargerel}),
these are:
\beq
C1_q \ : \qquad
q_d > 0 \quad (\Rightarrow q_u > 0)
\label{c1q}
\eeq
i.e., $Y_{Q_L} > 1$, which occurs if and only if $Y_{{\cal L}_L} < -N_c$, that
is,
\beq
q_e < -\biggl ( \frac{N_c+1}{2} \biggr )
\label{c1qe}
\eeq
\beq
C2_q \ : \quad q_u > 0 \ , \quad q_d < 0
\label{c2q}
\eeq
or equivalently, $-1 < q_d < 0$, which occurs iff
\beq
 - \biggl ( \frac{N_c+1}{2} \biggr ) < q_e < \biggl ( \frac{N_c-1}{2} \biggr )
\label{c2qe}
\eeq
and
\beq
C3_q \ : \quad q_u < 0 \quad (\Rightarrow q_d < 0)
\label{c3q}
\eeq
or equivalently, $q_d < -1$, which occurs iff
\beq
q_e > \biggl ( \frac{N_c-1}{2} \biggr )
\label{c3qe}
\eeq
A symmetric special charge within case $C2_q$ is
\beq
C2_{q,sym}: \ q_u = - q_d = \frac{1}{2} \quad \Longleftrightarrow \quad
q_\nu=-q_e=\frac{1}{2}
\label{c2qsym}
\eeq
Finally, there are two special cases which are borderline between $C1_q$ and
$C2_2$, and $C2_q$ and $C3_q$, respectively, and in which $q_u$ or $q_d$ is
electrically neutral:
\beq
C4_q \ : \quad q_d=0 \ , \quad q_u=1 \quad \Longleftrightarrow \quad
q_e = -\biggl ( \frac{N_c+1}{2} \biggr )
\label{qdzero}
\eeq
and
\beq
C5_q \ : \quad q_u=0 \ , \quad q_d=-1 \quad \Longleftrightarrow \quad
q_e = \biggl ( \frac{N_c-1}{2} \biggr )
\label{quzero}
\eeq
These cases are summarized in Table 1:

\begin{table}
\begin{center}
\begin{tabular}{|c|c|c|c|c|} \hline \hline & & & & \\
case & $q_d$ & $(q_u,q_d)$ & $Y_{Q_L}$ & $Y_{{\cal L}_L}$ \\
& & & & \\
\hline \hline
$C1_q$  & $> 0$          & $(+,+)$ & $> 1$ & $< -N_c$ \\ \hline
$C2_q$  & $-1 < q_d < 0$ & $(+,-)$ & $-1 < Y_{Q_L} < 1$ & $-N_c < Y_{{\cal
                                                    L}_L} < N_c$ \\ \hline
$C2_{q,sym}$ & $-1/2$ & $(1/2,-1/2)$  & 0 & 0 \\ \hline
$C3_q$  & $< -1$     & $(-,-)$ & $< -1$ & $> N_c$ \\ \hline
$C4_q$ & 0    & (1,0)    & 1 & $-N_c$ \\ \hline
$C5_q$ & $-1$ & $(0,-1)$ & $-1$ & $N_c$ \\ \hline
\hline
\end{tabular}
\end{center}
\caption{Possibilities for quark charges}
\label{table1}
\end{table}

 From eq. (\ref{yeq}), it is clear that
$q_u$ and $q_d$ are monotonically increasing (decreasing)
functions of $N_c$ if $q_e < -1/2$ ($q_e > -1/2$).  In the borderline case
$q_e=-1/2$, $q_u$ and $q_d$ are independent of $N_c$ (and equal to the
respective values in $C2_{q,sym}$), so that the anomalies of type (i) cancel
separately in the quark and lepton sectors.   For $N_c=3$, the
explicit conditions on $q_e$ for the five cases are: (1) $q_e < -2$; (2) $-2 <
q_e < 1$; (3) $q_e > 1$; (4) $q_e=-2$; and (5) $q_e=1$.  As these results show,
even for $N_c=3$, in the standard model with right-handed
components for all fields, the cancellation of anomalies does not imply that
any field, and in particular, any leptonic field, must have zero electric
charge.

\section{Classification of Solutions for Lepton Charges}

   The corresponding possible cases for leptonic ($\ell$) electric charges
are as
follows, taking $q_d$ as the independent variable in eq. (\ref{chargerel}):
\beq
C1_{\ell} \ : \quad
q_e > 0 \quad (\Rightarrow q_\nu > 0)
\label{c1e}
\eeq
iff
\beq
q_d < -\frac{1}{2}\biggl ( 1 + \frac{1}{N_c} \biggr )
\label{c1eqd}
\eeq

\beq
C2_\ell \ : \quad q_\nu > 0 \ , \quad q_e < 0
\label{c2e}
\eeq
iff
\beq
-\frac{1}{2} \biggl (1+\frac{1}{N_c} \biggr ) < q_d <
-\frac{1}{2}\biggl (1-\frac{1}{N_c} \biggr )
\label{c2eqd}
\eeq
\beq
C3_\ell \ : \quad q_\nu < 0 \quad (\Rightarrow q_e < 0)
\label{c3e}
\eeq
iff
\beq
q_d > -\frac{1}{2}\biggl (1-\frac{1}{N_c} \biggr )
\label{c3eqd}
\eeq
The symmetric subcase $C2_{\ell,sym}$ is identical to $C2_{q,sym}$ in
eq. (\ref{c2qsym}). The two special cases which
are borderline between $C1_\ell$ and $C2_\ell$, and between $C2_\ell$ and
$C3_\ell$ are, respectively
\beq
C4_\ell \ : \quad q_e=0 \ , \quad q_\nu=1 \quad \Longleftrightarrow \quad
q_d = -\frac{1}{2}\biggl (1+\frac{1}{N_c} \biggr )
\label{qezero}
\eeq
and
\beq
C5_\ell \ : \quad q_\nu=0 \ , \quad q_e=-1 \quad \Longleftrightarrow \quad
q_d = -\frac{1}{2} \biggl (1-\frac{1}{N_c} \biggr )
\label{qnuzero}
\eeq
These are summarized in Table 2:

\begin{table}
\begin{center}
\begin{tabular}{|c|c|c|c|c|} \hline \hline & & & & \\
case & $q_e$ & $(q_\nu,q_e)$ & $Y_{{\cal L}_L}$ & $Y_{Q_L}$ \\
& & & & \\
\hline \hline
$C1_\ell$  & $> 0$          & $(+,+)$ & $> 1$ & $< -1/N_c$ \\ \hline
$C2_\ell$  & $-1 < q_e < 0$ & $(+,-)$ & $-1 < Y_{{\cal L}_L} < 1$ &
                 $-1/N_c < Y_{Q_L} < 1/N_c$ \\ \hline
$C2_{\ell,sym}$ & $-1/2$ & $(1/2,-1/2)$  & 0 & 0 \\ \hline
$C3_\ell$  & $< -1$     & $(-,-)$ & $< -1$ & $> 1/N_c$ \\ \hline
$C4_\ell$ & 0    & (1,0)    & 1 & $-1/N_c$ \\ \hline
$C5_\ell$ & $-1$ & $(0,-1)$ & $-1$ & $1/N_c$ \\ \hline
\hline
\end{tabular}
\end{center}
\caption{Possibilities for lepton charges}
\label{table2}
\end{table}

Several comments are in order. First, note that $q_e$ and $q_\nu$ are
monotonically increasing (decreasing) functions of $N_c$ if $q_d < -1/2$
($q_d > -1/2$).  The special case $q_d=-1/2$ has been discussed above.
Secondly, observe that, even if we include right-handed neutrinos, so that
$q_\nu$ need not be zero in general, there is, for a given $N_c$, a
solution of (\ref{chargerel}) where it is zero, namely case
$C5_\ell$.

   Of the various cases of lepton charges, two would yield a world similar to
our own, in the sense that there would be neutral leptons with masses which are
naturally much less than the electroweak symmetry breaking scale $v$.  The
closest would be case $C5_\ell$, where the neutrino has zero charge.  As will
be discussed further below, case $C4_\ell$, with $q_e=0$ would also be
reminiscent of our world.  The lightness of the masses of the observed
electron-type leptons in this case would follow from a seesaw mechanism
completely analogous to that for the neutrinos in the physical world; in this
case, since $Y_{e_R}=0$, there would be gauge-invariant right-handed Majorana
mass terms of the form $\sum_{i,j=1}^{N_{gen.}} m_{R,ij}e_{iR}^T C e_{jR} +
h.c.$ in addition to the usual Dirac neutrino mass terms resulting from eq.
(\ref{yuk}).  By the usual argument, since the
right-handed electron Majorana mass terms are electroweak singlets, the mass
coefficients $m_{R,ij}$ are naturally much larger than the electroweak scale.
Diagonalizing the combined Dirac-Majorana
neutrino mass matrix would yield two sets of mass eigenvalues and corresponding
(generically Majorana) mass eigenstates, the observed, light electron-type
leptons having masses $\sim m_D^2/m_R << m_D$ and the heavy ones having
masses $\sim m_R$ (where $m_D$ denotes a generic Dirac mass, and we suppress
suppress generational indices).

 Although the generic situation in our generalization of the standard model
is that the electric charges of all the fundamental fermions are nonzero, there
are evidently four special cases in which one type of fermion has zero charge,
viz., $C4_q$ ($q_d=0$), $C5_q$ ($q_u=0$), $C4_\ell$ ($q_e=0$), and
$C5_\ell$ ($q_\nu=0$).  In each of the two leptonic cases containing a neutral
lepton, one may define a new model in which one excludes the right-handed Weyl
component for all generational copies of this lepton, viz., $e_{iR}$ for
case $C4_\ell$, and $\nu_{iR}$ for case $C5_\ell$, where $i=1.,,,N_{gen,}$.
Performing the excision of the $\nu_{iR}$ in case $C5_\ell$ and putting
$N_c=3$ just yields the standard model.  Performing the analogous excision of
the $e_{iR}$ fields in case $C4_\ell$ yields a model in which the electron-type
leptons are naturally light, for the same reason that the neutrinos are
naturally light in the standard model, namely that (a) there are no dimension-4
Yukawa terms contributing to the masses of electron-type leptons; and
(b) higher-dimension operators (which one would take account of when one views
the model as a low-energy effective field theory) give naturally small masses.
Indeed, the argument for the lightness of the neutral lepton
in these reduced models, $C4_\ell$ with no $e_{iR}$ fields, and $C5_\ell$ with
no $\nu_{iR}$ fields, could be regarded as more economical than the seesaw
mechanism, since the same result is achieved with a smaller field content
(albeit by making reference to higher-dimension, nonrenormalizable operators).
In passing, we note that in cases $C4_q$ and $C5_q$ where the conditions that
there be no $SU(2)^2U(1)_Y$ or $U(1)_Y^3$ anomalies, and no mixed
gauge-gravitational or global SU(2) anomalies, by themselves, would allow one
to define reduced models without $d_{iR}$ and $u_{iR}$ components,
respectively (analogously to the removal of $e_{iR}$ and $\nu_{iR}$ in the
leptonic cases $C4_\ell$ and $C5_\ell$), this is, of course, forbidden because
it would produce $SU(N_c)^3$ and $SU(N_c)^2U(1)_Y$ anomalies, as well as
rendering the color group chiral and thereby contradicting the observed absence
of parity and charge conjugation violation in strong interactions.

    An important observation concerns a connection between the values of
the lepton charges and the perturbative nature of the electroweak sector. In
the standard model, the observed electroweak decays and reactions are
perturbatively calculable.  However, in the generalized
$N_c$-extended standard model that we consider here, this is no longer
guaranteed to be the case, even if the
$SU(2)$ and $U(1)_Y$ gauge couplings $g$ and $g'$, and hence also the
electromagnetic coupling, $e=gg'/\sqrt{g^2+g'^2}=g\sin \theta_W$, are small.
Because the left-handed fermions have fixed, finite
values of weak $T_3=\pm 1/2$, the SU(2) gauge interactions are still
perturbative, as in the usual standard model.
However, for a given value of $N_c$, the solution to the anomaly
condition (\ref{yeq}) allows arbitrarily large values of the magnitudes of
fermion hypercharges and equivalently, electric charges, as is clear from the
explicit solutions (\ref{qdsol})-(\ref{qnusol}).  If $|q_d| >> 1$ (which,
for a fixed value of $N_c$, implies $|q_e| >> 1$), then even though the gauge
coupling $g'$ is small, the hypercharge interactions would involve strong
coupling, since $|g'Y_f| >> 1$ for each matter fermion $f$; similarly,
even though $g$ and hence $e$ are also small, the electromagnetic
interactions would also involve strong coupling, since $|eq_f| >> 1$ for each
matter fermion $f$.  Thus, nothing in the general
$N_c$-extended standard model (with right-handed components for all fermions)
guarantees that hypercharge and electromagnetic interactions are
perturbative, as observed in nature.  This perturbativity is natural (provided
that the
$g$ and $g'$ are small) only if one has a criterion for restricting the fermion
charges to values which are not $>> 1$ in magnitude.  Of course, this is
automatic in an approach using grand unification; here we inquire what
conditions make it natural without invoking grand unification.
There are only two cases where one can naturally guarantee that the fermion
charges are not $>> 1$ in magnitude (and these both
yield worlds  reminiscent of our own), namely $C4_\ell$ and $C5_\ell$,
where $q_e=0$ or $q_\nu=0$, and the electron-type leptons and neutrinos,
respectively, are
naturally very light compared to the electroweak scale.  In these cases, as
eqs. (\ref{qezero}) and (\ref{qnuzero}) show, the quark charges cannot be
large in magnitude.  Of course, any set of charges in which
$|q_e|$ (and hence $|q_\nu|$)
are bounded above by a number of order unity implies by
(\ref{chargerel}) that $|q_d|$ and $|q_u|$ are
also bounded above in magnitude by $O(1)$, but one would lack a
specific reason for choosing such a value of $q_e$ or $q_\nu$.  We thus are led
to conclude that, in the context of the general $N_c$-extended standard model,
the condition that there be neutral (electron- or neutrino-type) leptons
which are much lighter than the electroweak scale provides a natural way to get
fermion charges which are not $>> 1$ in magnitude and hence to get perturbative
hypercharge and electromagnetic interactions, given that the electroweak gauge
couplings are small.   Note that this is true both in cases $C4_\ell$ and
$C5_\ell$ themselves and in the reduced models in which one
excludes the right-handed components of the
respective neutral leptons, $e_{iR}$ in $C4_\ell$ and
$\nu_{iR}$ in $C5_\ell$, since in either case, albeit for
different reasons (seesaw mechanism or higher-dimension operators), one has
naturally light neutral leptons.

\section{Conditions for Finiteness of Electroweak Effects as $N_c \to
\infty$}

    We have already noted that the anomaly conditions can be solved for fermion
hypercharges and equivalently electric charges of arbitrarily large
magnitude. Obviously, one condition for hypercharge and electromagnetic
interactions to be finite is that one choose finite values of fermion charges
to solve eq. (\ref{chargerel}).  It is also of interest to consider this from
the viewpoint of the large-$N_c$ limit.
 From eqs. (\ref{qesol}) and (\ref{qnusol}) it is clear that if $Y_{Q_L}$ is
nonzero, then $q_e$ and $q_\nu$ will diverge, like $(-1/2)Y_{Q_L}N_c$, as $N_c
\to \infty$.  A necessary condition for the lepton charges to remain finite in
this limit is that
\beq
\lim_{N_c \to \infty} q_d = -\frac{1}{2}
\label{qdlim}
\eeq
i.e., $\lim_{N_c \to \infty}Y_{Q_L}=0$.
However, this is not a sufficient condition; for example,
if, as a function of $N_c$, $q_d$ behaves as
\beq
q_d \to \frac{-1+aN_c^{-\alpha}}{2}
\label{qdexample}
\eeq
for large $N_c$ (where $a \ne 0$), then, from eq. (\ref{qesol}),
\beq
q_e = -\frac{1}{2}(1 + aN_c^{1-\alpha})
\label{qeexample}
\eeq
which is finite as $N_c \to \infty$ if and only if $\alpha \ge 1$.  In
contrast, as is clear from (\ref{qdsol}), for any fixed (finite) value of
$q_e$, $q_d$ has a finite limit, namely $q_d=-1/2$, as $N_c \to \infty$.

   However, this is still not sufficient for electroweak effects to remain
finite in the limit $N_c \to \infty$.
It will be recalled that in the large--$N_c$ limit, one holds
\beq
g_s^2 N_c = const.
\label{g3condition}
\eeq
where $g_s$ denotes the $SU(N_c)$ gauge coupling
\cite{thooft}-\cite{wittennc}. As has
been noted in Ref. \cite{cy}, to avoid a breakdown of large--$N_c$ relations
such as that for the $\pi^0 \to \gamma\gamma$ amplitude while retaining nonzero
electroweak interactions as $N_c \to \infty$, one sets
\beq
g^2N_c = const.
\label{grel}
\eeq
and
\beq
(g')^2 N_c = const.
\label{gprimerel}
\eeq
in this limit,
where $g$ and $g'$ denote the SU(2) and $U(1)_Y$ gauge couplings, and the
constants in eqs. (\ref{g3condition}), (\ref{grel}) and (\ref{gprimerel}) are,
of course, different. It is easily
seen that this is true for our generalization with right-handed neutrino fields
and variable lepton charges, just as it was true of the generalization
considered in Ref. \cite{cy} without any $\nu_{iR}$ fields and with fixed,
conventional lepton charges.  Hence also, the electromagnetic coupling
$e=gg'/\sqrt{g^2+g'^2}$ satisfies the same scaling property
\beq
e^2N_c = const.
\label{emcon}
\eeq
as $N_c \to \infty$.

\section{Relations Between Quark and Lepton Charge Classes}

   It is of interest to work out the relationships between the various cases
describing the possible quark and leptons charges.
We thus consider a value of $q_d$ lying in a given class, $C1_q$
through $C5_q$, and determine to which class the corresponding lepton charges
determined by eq. (\ref{chargerel}) belong. First, as one can see from Table 1,
the
condition that the quark charges fall in class $C1_q$ implies that the lepton
charges fall in class $C3_\ell$.  We symbolize this as
\beq
q_d \in C1_q \quad \Longrightarrow \quad q_e \in C3_\ell
\label{c1qimplic}
\eeq
The converse does not, in general, hold.
The other implications are listed below (and again, the converses do not, in
general, hold, except for $C2_{q,sym}$):
\beq
q_d \in C2_{q,sym} \quad \Longleftrightarrow \quad q_e \in C2_{\ell, sym}
\label{c2qsymimplic}
\eeq
\beq
q_d \in C3_q  \ \ or \ \ C5_q\quad \Longrightarrow \quad q_e \in C1_\ell
\label{c3qimplic}
\eeq
\beq
q_d \in C4_q \quad \Longrightarrow \quad q_e \in C3_\ell
\label{c4qimplic}
\eeq
The condition that $q_d \in C2_q$ can be met for certain values of $q_e$ in
each of the leptonic charge classes.
The implications following from a given leptonic charge class are
\beq
q_e \in C1_\ell \quad \Longrightarrow \quad q_d \in C2_q \ , \ C3_q \ ,
\ \ or \ \ C5_q
\label{c1eimplic}
\eeq
\beq
q_e \in C2_\ell \quad \Longrightarrow \quad q_d \in C2_q
\label{c2eimplic}
\eeq
\beq
q_e \in C3_\ell \quad \Longrightarrow \quad q_d \in
C1_q \ , \ C2_q \ , \ \  or \ \ C4_q
\label{c3eimplic}
\eeq
\beq
q_e \in C4_\ell \ \ or \ \ C5_\ell \quad \Longrightarrow
\quad q_d \in C2_q
\label{c4eimplic}
\eeq

\section{A Relation Connecting Certain Pairs of Solutions}

   Two respective solutions $S$ and $S'$ of (\ref{chargerel})
with $q_d$ (and a resultant $q_e$) and $q_d'$ (and a resultant
$q_e'$) have a certain simple relation if the corresponding hypercharges
satisfy
\beq
Y_{Q_L} = - Y_{Q_L}'
\label{yqrel}
\eeq
or equivalently, by eq. (\ref{chargerel}),
\beq
Y_{{\cal L}L} = - Y_{{\cal L}_L}'
\label{ylrel}
\eeq
In terms of the fermion charges, these equivalent conditions read
\beq
q_d + q_d' + 1 = 0
\label{qdrel}
\eeq
i.e.,
\beq
q_e + q_e' + 1 = 0
\label{qerel}
\eeq
To see the relation, we recall that the constraint (\ref{yeq}) implies that
all of the hypercharges for the two cases can be expressed in terms of any one,
say $Y_{Q_L}$ and $Y_{Q_L}'$, respectively; further, one can use the condition
(\ref{yqrel}) to express all hypercharges in terms of $Y_{Q_L}$.  Then the
fields for the original solution $S$ are
\beq
Q_L \ : \ (N_c,2,Y_{Q_L})
\label{qlexp}
\eeq
\beq
u_R \ : \ (N_c,1,1+Y_{Q_L})
\label{urexp}
\eeq
\beq
d_R \ : \ (N_c,1,-1+Y_{Q_L})
\label{drexp}
\eeq
\beq
{\cal L}_L \ : \ (1,2,-N_cY_{Q_L})
\label{llexp}
\eeq
\beq
\nu_R \ : \ (1,1,1-N_cY_{Q_L})
\label{nurexp}
\eeq
\beq
e_R \ : \ (1,1,-1-N_cY_{Q_L})
\label{erexp}
\eeq
Now, expressing the field content of solution $S'$ in terms of the
charge-conjugates fields,
\beq
(Q^c_R)' \ : \ (N_c^*,2,Y_{Q_L})
\label{qcrprime}
\eeq
\beq
(u^c_L)' \ : \ (N_c^*,1,-1+Y_{Q_L})
\label{uclprime}
\eeq
\beq
(d^c_L)' \ : \ (N_c^*,1,1+Y_{Q_L})
\label{dclprime}
\eeq
\beq
({\cal L}^c_R)' \ : \ (1,2,-N_cY_{Q_L})
\label{lcrprime}
\eeq
\beq
(\nu^c_L)' \ : \ (1,1,-1-N_cY_{Q_L})
\label{nuclprime}
\eeq
\beq
(e^c_L)' \ : \ (1,1,1-N_cY_{Q_L})
\label{eclprime}
\eeq
where our notational convention is $\psi^c_R \equiv ((\psi_L)^c)_R$,
$\psi^c_L \equiv ((\psi_R)^c)_L$.
Evidently, there is a one-to-one correspondence between the fields
(\ref{qlexp}) -- (\ref{erexp}) of solution $S$ and the fields
(\ref{qcrprime}) -- (\ref{eclprime}) of solution $S'$ according to which $L
\leftrightarrow R$, $N_c \to N_c^*$ (i.e., fundamental
reprepresentation is replaced by conjugate fundamental representation of the
$SU(N_c)$ color group), and $(u^c_L)' \to d_R$, $(d^c_L)' \to u_R$,
$(\nu^c_L)' \to e_R$ and $(e^c_L)' \to \nu_R$, etc.
(Here we use the fact that the
representations of SU(2) are (pseudo)real.)
In particular, the leptonic fields ${\cal L}_L$, $\nu_R$, and $e_R$ for
solution $S$ transform according to precisely the same representations of
$SU(N_c) \times SU(2) \times U(1)_Y$ as the lepton fields $({\cal L}^c_R)'$,
$(e^c_L)'$, and $(\nu^c_L)'$ of solution $S'$, respectively.  We note that the
special cases
describing the quark charges (and their corresponding lepton charges) in $C4_q$
and $C5_q$ satisfy the condition (\ref{yqrel}) (and the equivalent
eq. (\ref{ylrel}) for the leptons), so that $(C4_q,C5_q)$ form such a pair
$(S,S')$ of
solutions.  Similarly, the lepton charges (and their corresponding quark
charges) in $C4_\ell$ and $C5_\ell$ satisfy the condition (\ref{ylrel})
(and the equivalent eq. (\ref{yqrel}) for the quarks), so that
$(C4_\ell,C5_\ell)$ form another such pair $(S,S')$.  There is a (continuous)
infinity of other pairs of solutions forming such pairs with hypercharges which
are equal and opposite.  Finally, the symmetric case $C2_{q,sym} =
C2_{\ell,sym}$ with $Y_{Q_L}=0=Y_{{\cal L}_L}$ also satisfies the condition
(\ref{yqrel}) and thus forms a pair with itself $(S,S'=S)$.

\section{Some Properties of Various Cases}

   We next comment on some properties of various classes of solutions of
eq. (\ref{chargerel}).  In this discussion, we consider both fixed, finite
$N_c$ and the limit $N_c \to \infty$.  The hadronic
spectrum of the theory would contain
various meson and glueball states, the latter being, in general, mixed with
$\bar q q$ mesons of the same quantum numbers to form physical mass
eigenstates.  Independent of the specific values of fermion charges, the
$\bar q q$ meson charges would always be 0 or $\pm 1$ (the latter because
$q_u=q_d+1$).
Other aspects of the spectrum would depend on the nature of
electroweak symmetry breaking, such as superpartners in the MSSM; we shall not
discuss these here.  There are some interesting general results which one can
derive concerning baryons, and we proceed to these.

\subsection{Baryons}

We consider baryons composed of $r$ up-type and $N_c-r$ down-type quarks
\cite{alt} and
denote their electric charge as $q(B(r,N_c-r))$.   (For considerations of
electric charge, one can suppress the flavor dependence of the quark
constituents; thus, for example, by down-type quarks, we include $d$, $s$,
and $b$).  The electric charge of the baryon(s) $B(r,N_c-r)$ is
\beqs
q\Bigl ( B(r,N_c-r) \Bigr ) & = & r + N_cq_d \nonumber \\
                          & = & r - \frac{1}{2}(N_c + 2q_e+1)
\label{qbgeneral}
\eeqs
In the special case where all of the
up-type quarks and down-type are $u$ and $d$, respectively, two baryons which
are related by a strong-isospin rotation $u \leftrightarrow d$ are
$B(r,N_c-r)$ and $B(N_c-r,r)$.
The charge difference between these is
\beq
q \Bigl ( B(r,N_c-r)\Bigr ) - q\Bigl ( B(N_c-r,r)\Bigr ) = 2r-N_c
\label{mirrorchargedif}
\eeq
Now we assume that $m_u, m_d << \Lambda_{QCD}$ (where $\Lambda_{QCD}$ is the
scale characterizing the $SU(N_c)$ color interactions), as in the physical
world.  A strong--isospin mirror pair which constitutes a kind of
generalization of the proton and neutron is the (light $u$, $d$ quark, spin
1/2) pair
\beq
{\cal P} = B \biggl ( \frac{N_c+1}{2}, \frac{N_c-1}{2} \biggr )
\label{pp}
\eeq
and
\beq
{\cal N} =B \biggl ( \frac{N_c-1}{2}, \frac{N_c+1}{2} \biggr )
\label{nn}
\eeq
For $N_c=3$, ${\cal P} = p$, ${\cal N} = n$.
 From eq. (\ref{qbgeneral}), it follows that
\beq
q_{\cal P} = -q_e
\label{qp}
\eeq
and
\beq
q_{\cal N}=-q_\nu
\label{qn}
\eeq
(and furthermore $q_{\cal P} = q_{\cal N}+1$).
Some further general results on baryon charges are the following.  First, if
and only
if $q_d$ and $q_u$ have the same sign, as they do for cases $C1_q$ and $C3_q$,
then all baryons also have the same sign of electric charge.   Second, as is
clear from eqs. (\ref{qp}) and (\ref{qn}), the proton and neutron ${\cal P}$
and ${\cal N}$ have the same sign of electric charge if and only if $q_e$ and
$q_\nu$ have the same sign, which holds in cases $C1_\ell$ and $C3_\ell$.  Thus
also, $sgn(q_{\cal P}) = -sgn(q_{\cal N}) \Longleftrightarrow sgn(q_e) =
-sgn(q_\nu) \Longleftrightarrow q_e \in C2_\ell$.  The special case $C5_\ell$,
with $q_\nu=0$ yields proton and neutron charges the same as in our world,
while in the special case $C4_\ell$, with $q_e=0$, one would have $q_{\cal
P}=0$ and $q_{\cal N}=-1$.  In case $C2_{q,sym}$, $q_{\cal P} =
-q_{\cal N}=1/2$.
In the cases $C1_\ell$ and $C3_\ell$ in which ${\cal P}$ and ${\cal N}$
have the same sign of electric charge, the resultant Coulomb replusion would
increase the energy (i.e. decrease the binding energy) of a nucleus, as
compared to the cases where these nucleons have opposite signs of electric
charge or one is neutral.  Consequently, in these cases the Coulomb
interaction could destabilize certain nuclei which are stable in the physical
world.

\subsection{Atoms}

   As a consequence of the charge relation eq. (\ref{qp}), for all cases of
fermion charges (satisfying (\ref{chargerel})) except the case
$C4_\ell$ ($q_e=0$), there will exist a Coulomb bound state of the
proton ${\cal P}$ and electron, which is the $N_c$-extended
generalization of the hydrogen atom.
Furthermore, for all cases of fermion charges
(satisfying eq. (\ref{chargerel})) except $C5_\ell$ ($q_\nu=0$), there will
exist a second neutral Coulomb bound state, which has no analogue in the
usual $N_c=3$ standard model, namely, $({\cal N}\nu_{r=1})$,
where $\nu_{r=1}$ denotes the lightest neutrino mass eigenstate.
Since the standard model conserves baryon number
perturbatively\cite{npert}, it follows that, if $q_e \ne 0$, so that the
generalized H atom, $({\cal P}e)$, exists, this bound state is stable.  Even
for cases other than $C5_\ell$, where the $({\cal N} \nu_{r=1})$ atom
exists, it would decay weakly, as a consequence of the decay
${\cal N} \to {\cal P} + e + \bar\nu_e$ (this
really means the decays ${\cal N} \to {\cal P} + e + \bar\nu_r$, involving all
mass eigenstates $\nu_r$ in the weak eigenstate $\nu_e = \sum_{r=1}^3
U_{1r}\nu_r$ which are kinematically allowed to occur in the final state).
For nonzero $q_e$ and $q_\nu$, no leptons would be generically
expected to be very light compared with other fermions.  However, depending on
the fermion mass spectrum, there could exist other stable Coulomb bound
states.  For example, assuming the usual electron-type lepton mass spectrum,
if $m(\nu_{r=2}) < 2m_e + m(\nu_{r=1})$, then the state $({\cal N}\nu_{r=2})$
could also be stable.  Henceforth, among the possible $({\cal N}\nu_{r})$
states, we shall only consider $({\cal N}\nu_{r=1})$ and shall suppress
the $r=1$ subscript in the notation.

      In the following discussion of the H and $({\cal N}\nu)$ atoms,
we shall implicitly assume, respectively, that $q_e \ne 0$ and $q_\nu \ne 0$,
so that these atoms exist.
The condition that the H atom is a nonrelativistic bound state is
\beq
({\cal P}e) \ \ nonrel. \ \ \Longleftrightarrow |q_e| \alpha << 1
\label{nonrelh}
\eeq
Given that $q_\nu=q_e+1$, this is effectively the same condition for the
$({\cal N}\nu)$ bound state:
\beq
({\cal N}\nu) \ \ nonrel. \ \ \Longleftrightarrow |q_\nu| \alpha << 1
\label{nonreln}
\eeq
Note that if this condition is not met, i.e., if $|q_e|\alpha \gtwid 1$, then
the electromagnetic interaction between ${\cal P}$ and $e$ would involve
strong coupling. Assuming that these states are nonrelativistic, the
binding energy of the ground state of the H atom is
given, to lowest order, by
\beq
E_{({\cal P}e)} = -\frac{(q_e\alpha)^2m_{red.}}{2}
\label{pebinding}
\eeq
where $\alpha=e^2/(4\pi)$ and the reduced mass is
$m_{red.}=m_{\cal P}m_e/(m_{\cal P}+m_e)$.  The Bohr radius for this ground
state of the H atom would be
\beq
a_0 = \frac{1}{q_e^2\alpha m_{red.}}
\label{a0}
\eeq
Formulas (\ref{pebinding}) and (\ref{a0}) apply to the
$({\cal N}\nu)$ bound state with the obvious replacements ${\cal P} \to
{\cal N}$ and $e \to \nu$.

   For a fixed value of $q_e$ and hence $q_\nu$, such that the respective bound
state $({\cal P}e)$ or $({\cal N}\nu)$ exists, assuming that the lepton
masses are fixed, the magnitudes of the
respective binding energies decrease as $N_c$ increases.
To see this, note first that since $m_{\cal P},
m_{\cal N} \sim N_c$ in this limit \cite{wittennc}, the respective
reduced mass $m_{red} \to m_e$ or
$m_\nu$ and hence is finite.  Then from eq. (\ref{emcon}), it follows
that
\beq
E_{({\cal P}e)} \ , E_{({\cal N}\nu)} \sim N_c^{-2}
\label{vnc}
\eeq
as $N_c$ gets large.

   If both $q_e$ and $q_\nu$ are nonzero, then atoms with nuclei having
atomic number $A \ge 2$ would exihibit a
qualitatively new feature not present in our world: the leptons bound to the
nucleus would be of two different types.  A generic atom would be of the form
\beq
\Bigl ( nucl(N_{\cal P}{\cal P}, N_{\cal N}{\cal N}); N_{\cal P}e,
N_{\cal N}\nu \Bigr )
\label{genatom}
\eeq
Whereas the characteristic size of atoms and molecules in the
physical world is set by the Bohr radius, these atoms would have
charged lepton clouds characterized by different sizes, reflecting their
different masses and charges.  These higher-$A$ nuclei and atoms
would undergo weak decays via $e^-$ or $e^+$ emission or $e^-$ capture,
as in the physical world, and, in addition, $\nu$ or $\bar\nu$ emission or
$\nu$ capture.

\subsection{Possible Lepton-Lepton Coulomb Bound State}

    There could also occur a stable purely leptonic Coulombic bound
state, $(e \nu_{r=1})$. A necessary condition for this would be that
$q_e$ and $q_\nu$ are opposite in sign, i.e. that the lepton charges fall
in case $C2_\ell$.  However, unlike the $({\cal P}e)$ and $({\cal N}\nu_{r=1})$
atoms, this leptonic state (and other possible ones e.g. for $r=2$, would, in
general, have a nonzero charge, namely, $q\Bigl ( (e \nu) \Bigr ) =
Y_{{\cal L}_L}$.
In the physical world, one knows of many such Coulombic bound states with
net charge, such as the negative hydrogen ion $(pee) = H^-$.

\subsection{Lepton Masses for the Case $q_\nu=-q_e=1/2$}

  Clearly, $q((e \nu_r))=0$ only for the symmetric subcase $C2_{\ell,sym} =
C2_{q,sym}$ in eq. (\ref{c2qsym}), where $q_\nu=-q_e=1/2$ so that
$Y_{{\cal L}_L}=0$.  In this case, the lepton mass spectrum exhibits some
unusual features, which depend, moreover, on whether $N_{gen.}$ is even
or odd.  We note first that, in addition to the lepton Yukawa
interactions in (\ref{yuk}), there would two more such terms, so that the total
leptonic Yukawa part of the Lagrangian would be
\beqs
-{\cal L}_{Yuk,C2_{\ell,sym}} & = & \sum_{i,j=1}^{N_{gen.}} \biggl [
\Bigl ( Y^{(e)}_{ij}\bar {\cal L}_{ibL}e_{jR} +
   Y^{(\nu2)}_{ij}\epsilon_{ab}{\cal L}^{Ta}_{iL}C \nu^c_{jL} \Bigr )H^b_d
 \nonumber \\
                              & &
+\Bigl (Y^{(\nu)}_{ij}\bar {\cal L}_{ibL}\nu_{jR} +
    Y^{(e2)}_{ij}\epsilon_{ab}{\cal L}^{Ta}_{iL}C e^c_{jL} \Bigr )H^b_u
\biggr ] + h.c.
\label{lepyuk}
\eeqs
where $i,j$ are generation indices and $a,b$ are SU(2) indices, and
the notation applies to either the standard model or the MSSM, as discussed
before, following eq. (\ref{yuk}).
The vev's of the Higgs fields would give rise to the mass terms
\beq
-{\cal L}_{Yuk,mass} = \sum_{i,j=1}^{N_{gen.}} \biggl [
   M^{(e)}_{ij} \bar e_{iL}e_{jR} +M^{(\nu2)}_{ij} \bar\nu_{jR}\nu_{iL}
+ M^{(\nu)}_{ij} \bar\nu_{iL} \nu_{jR} -
  M^{(e2)}_{ij} \bar e_{jR}e_{iL} \biggr ] + h.c.
\label{massterms}
\eeq
where
\beq
M^{(e)}= 2^{-1/2}Y^{(e)}v_d \ , \quad M^{(\nu2)} =
                           2^{-1/2}Y^{(\nu2)}v_d
\label{mlower}
\eeq
\beq
M^{(\nu)}= 2^{-1/2}Y^{(\nu)}v_u \ , \quad M^{(e2)} =
                           2^{-1/2}Y^{(e2)}v_u
\label{mupper}
\eeq
and we have used $\nu_{iL}^T C \nu^c_{jL}=\bar\nu_{jR}\nu_{iL}$ and
$e_{iL}^T C e^c_{jL}=\bar e_{jR}e_{iL}$.  (In a theory without Higgs fields,
these mass terms would arise, as discussed before, from certain
multifermion operators.)
Furthermore, there would be the electroweak-singlet bare leptonic mass terms
\beq
-{\cal L}_{bare} = \sum_{i,j=1}^{N_{gen.}} \biggl [
M^{(L)}_{ij}\epsilon_{ab}{\cal L}_{i L}^{Ta} C {\cal L}^b_{_{j L}} +
M^{(R)}_{ij}e_{i R}^T C \nu_{_{j R}} \biggr ] + h.c.
\label{baremass}
\eeq
(Note that $M_{1,ij}$ is automatically antisymmetric; $M_{1,ij}=-M_{1,ji}$ and
that $\nu_{i R}^T C e_{_{j R}}=e_{jR}^TC \nu_{iR}$.)
The mass coefficients multiplying these electroweak-singlet terms would be
naturally much larger than the electroweak symmetry breaking scale $v$. Without
having to analyze the full set of mass terms in eqs. (\ref{massterms}) and
(\ref{baremass}) in detail, one can thus immediately conclude that the
SU(2)-singlet lepton fields will pick up masses which are naturally much larger
than the EWSB scale.  Indeed, (whether $N_{gen.}$
is even or odd) one can always rewrite the SU(2)-singlet leptons as
four-component Dirac fields; explicitly, these are

\beq
\psi_{i} = \left (\begin{array}{c}
                  \nu_{iR} \\
                   e^c_{iL} \end{array} \right ) \ , i = 1..., N_{gen.}
\label{psi}
\eeq
with charge $q_\psi=1/2$ (here the spinor refers to Dirac, not SU(2), space,
and we use a representation in which $\gamma_5$ is diagonal)
These form bare mass terms $\sum_{i=1}^{N_{gen.}}m_{Di}\bar\psi_i\psi_i$ with
masses $m_{D,i}$ naturally $>> v$.

In the hypothetical situation in which
$N_{gen.}$ is even, one could form $(N_{gen.}/2)$ Dirac SU(2) doublets in a
similar manner, combining ${\cal L}_{1L}$ and ${\cal L}^c_{2R}$ into the
first doublet, say, ${\cal L}_{3L}$ and ${\cal L}^c_{4R}$ into the second one,
and so forth for the others.  Here we use the fact that the representations of
SU(2) are (pseudo)real.  Again, these SU(2) doublets would form
electroweak-singlet Dirac bare mass terms with masses which are naturally much
larger than the EWSB scale.  Note that the only gauge interaction
of these Dirac doublets, namely that involving the SU(2) gauge fields, is
vectorial.  This rewriting of the doublets as Dirac fields coupling in a
vectorial manner is similar to the method that we used earlier in lattice gauge
theory studies \cite{su2lgt}.  The matter fermions in the effective field
theory at and below the electroweak scale then consist only of the quarks.
In this hypothetical case, since $N_{gen.}$ is
even, there are $N_{gen.}N_c$ doublets of quarks, which is even whether or not
$N_c$ is even or odd, so this effective field theory is free of any global
SU(2) anomaly.  (Recall that for the present case, $C2_{q,sym}=C2_{\ell,sym}$,
the anomalies of type (i) cancel individually for the quark and lepton
sectors.)

For the case where $N_{gen.}$ is odd,
one can form $(N_{gen.}-1)/2$ Dirac SU(2) doublets which naturally gain large
masses, as above.  To see what happens to the one remaining leptonic chiral
SU(2) doublet, it is sufficient to deal with a simple $N_{gen.}=1$
example. Using the fact that the masses multiplying the electroweak-singlet
bare mass terms are naturally much larger than the electroweak scale, while the
masses resulting from the Yukawa couplings are
$\ltwid v$, and diagonalizing the
mass matrix, one finds two large eigenvalues $\pm M_R$, and two very small
eigenvalues, $\pm m_1m_2/M_R$ where $M_R$, $m_1$, and $m_2$ denote the
coefficients of $e_R^TC\nu_R$, $\bar\nu_L\nu_R$, and $\bar e_L e_R$,
respectively.  The two large eigenvalues correspond, up to very small
admixtures, to the SU(2)-singlet states already discussed above.  The masses
gained by the components in the remaining SU(2)-doublet are naturally much
less than the electroweak scale, because of
a kind of seesaw mechanism.  The effective field theory (EFT) at and below
the electroweak scale would consist of a remainder of
$N_{d,EFT} = (N_{gen.}N_c+1)$ SU(2) doublets. Since $N_{gen.}$ and hence
$N_c$ are odd, $N_{d,EFT}$ is even, so again
there is no global SU(2) anomaly in this sector.  The sector containing
high-mass leptons was rewritten in vectorial form, and hence is obviously free
of any anomalies.

\section{Question of Grand Unification}

    Finally, we address the issue of grand unification of the $N_c$--extended
standard model with right-handed neutrinos.  Although in the standard model
(with no right-handed neutrinos) and its $N_c$-extension, one already gets
charge quantization without grand unification, the latter does provide an
appealingly simple
(if not unique \cite{str}) way to obtain gauge coupling unification.
Here we shall prove a strong negative result. Our proof
will essentially consist of a counting argument and will not make use of
the specific fermion charge assignments obtained as solutions to
eq. (\ref{chargerel}).  Our proof will apply both for the case of odd
$N_{gen.}$ and for the hypothetical case of even $N_{gen.}$.
Again, we shall not need to make any explicit
assumption concerning the nature of electroweak symmetry breaking.
Clearly, however, if one discusses grand unification at all, it is natural
to assume that physics remains perturbative from the electroweak scale up to
the scale of the grand unified theory (GUT) (so that the unification of gauge
couplings is not an accident)
and hence work within the framework of a supersymmetric
theory. This is also motivated by the fact that supersymmetry can protect the
Higgs sector against large radiative
corrections and, if the $\mu$ problem can be solved, can thereby account for
the gauge hierarchy in a GUT \cite{split}.
We follow the standard rules of grand unification: first, in
order have light fermions, one must use a group with complex representations.
Second, in order to have natural cancellation of anomalies in gauged currents,
one restricts the choices of a group to those which are ``safe'' (i.e., have
identically zero triangle anomaly for all fermion representations) \cite{gg}.
Note that at the GUT level, there are no mixed gauge-gravitational anomalies,
since those necessarily involve a U(1) gauge group, and also no global
anomaly involving the GUT group itself,
since $\pi_4(SU(N))=\emptyset$ for $N \ge 3$,
$\pi_4(SO(N))=\emptyset$ for $N \ge 6$, and $\pi_4(E_6)=\emptyset$
 \cite{hu,witten,rem} (of course, the restriction
(\ref{global}) still holds).  Now although $E_6$
has complex representations and is safe, it has a fixed rank of 6, and hence
cannot be used for general $N_c$.  The natural choice of GUT gauge group is
thus $SO(4k+2)$ with $k \ge 2$.  Now, as in the original discussion
\cite{gggut}, one must satisfy an inequality on ranks: for our case, in
order for $SU(N_c) \times SU(2) \times U(1)_Y$ to be embedded in a GUT group
$G$, it is necessary that
\beq
rank(G) \ge N_c+1
\label{rank}
\eeq
Using the standard result
\beq
rank(SO(2n))=n \ ,
\label{rankso}
\eeq
setting $2n=4k+2$, and substituting (\ref{rank}), we obtain the inequality
\beq
2k \ge N_c
\label{tknc}
\eeq
If $N_{gen.}$ is odd, and hence, by (\ref{global}), $N_c$ is odd, this becomes
\beq
2k \ge N_c + 1 \ \ for \ \ N_c \ \ odd
\label{2krel}
\eeq
We are thus led to consider the special orthogonal group
$G = SO(2N_c+4)$, (corresponding to the algebra
$D_{N_c+2}$, in the Cartan notation) with
rank $N_c+2$.  Since $N_c+2$ is odd, $G = SO(2N_c+4)$ has
a complex spinor representation of dimension $2^{N_c+1}$.
Now we would like to fit all of the fermions of each generation
in a Weyl field transforming according to the spinor representation
of this group.  Note that in order for this to be possible, it is necessary
that
\beq
\sum_{f} Y_f = 0
\label{ygenerator}
\eeq
for each generation, since $Y$ is now a generator of a simple (nonabelian)
group, and hence its trace must be zero.  This requirement is met automatically
as a consequence of the vectorial nature of the electromagnetic coupling; the
left-hand side of eq. (\ref{ygenerator}) is, indeed, identical to that of
eq. (\ref{ggu1}) discussed before. Now there are
\beq
N_f = 4(N_c+1)N_{gen}
\label{nf}
\eeq
Weyl matter fermion fields in the theory\cite{exc}.
Requiring that these fit precisely in $N_{gen.}$ copies of
the spinor representation then yields the condition
\beq
2^{N_c+1} = 4(N_c+1)
\label{guteq}
\eeq
But this has a solution only for $N_c=3$.  This is a very interesting
result, and perhaps gives us a deeper understanding of why $N_c=3$ in our
world.

  We would reach the same conclusion even if $N_{gen.}$ were even.  In this
case, (\ref{global}) allows $N_c$ to be either even or odd.  If $N_c$ is odd,
the same reasoning as before applies directly.  If $N_c$ is even, then
(\ref{tknc}) can be satisfied as an equality, so that the group would be
$SO(2N_c+2)$ (with minimal rank $N_c+1$) corresponding to $D_{N_c+1}$,
rather than $SO(2N_c+4)$.  Now since
$N_c$ is even, the dimension of the spinor representation of $SO(2N_c+2)$ is
$2^{N_c+1}$.  Hence, we are led to the same condition, (\ref{guteq}) as
before, and the same conclusion follows.

\section{Concluding Remarks}

    In summary, we have explored the implications of the cancellation of
anomalies for the $N_c$--extended standard model with right-handed components
for all fermions. We have shown that anomaly cancellation does not imply the
quantization of the fermion charges and have discussed
some interesting properties of various classes of solutions for these
charges. In particular, we have related the condition that there be neutral
leptons with masses much less than the electroweak scale to the feature that
the fermion charges are not $>> 1$ in magnitude and hence that the electroweak
interactions are perturbative.  Finally, we have proved that the unification
of the $SU(N_c) \times SU(2) \times U(1)_Y$ theory in $SO(2N_c+4)$
(with the usual assignment of the fermions to the spinor representation)
can only be carried out for $N_c=3$.
The world which we analyze here is, of course, a
generalization of our physical one, but we believe that, as with the original
$1/N_c$ expansion in QCD, by thinking about the standard model in
a more general context, one may gain a deeper understanding of its features.

\vspace{6mm}

   The author acknowledges with pleasure the stimulating influence of Ref.
\cite{cy} and a conversation with Prof. Tung-Mow Yan in prodding him to write
up these results.
The current research was supported in part by the NSF grant PHY-93-09888.

\vfill
\eject

\vspace{6mm}

\begin{thebibliography}{99}

\bibitem{sm}{S. L. Glashow, Nucl. Phys. {\bf B22}, 579 (1961);
S. Weinberg, Phys. Rev. Lett. {\bf 27}, 1264 (1967);
A. Salam, in {\it Elementary Particle Theory}, Proceedings of the 8th Nobel
Symposium, ed. N. Svartholm (Almqvist and Wiksell, Stockholm, 1968);
S. L. Glashow, J. Iliopoulos, and L. Maiani, Phys. Rev. {\bf D2}, 1285 (1970).}

\bibitem{kk}{Indeed, as is well known, the effort to derive known gauge
interactions from a deeper theory involving gravity has a long history dating
back at least to the works of Kaluza and Klein in 1921 and 1926.}

\bibitem{thooft}{G. 't Hooft, Nucl. Phys. {\bf B72}, 461 (1974); {\it ibid},
{\bf B75}, 461 (1974).}

\bibitem{nclit}{There is a considerable literature on large-$N_c$ methods.
An early review is Ref. \cite{coleman}.  For a discussion of the $1/N_c$
expansion applied to 4D QCD, see Ref. \cite{wittennc}.}

\bibitem{coleman}{S. Coleman, Lectures at the 1979 Erice Summer School, in {\it
Pointlike Structure Inside and Outside Hadrons}, ed. A. Zichichi (Plenum,
1982), p. 11.}

\bibitem{wittennc}{E. Witten, Nucl. Phys. {\bf B160}, 57 (1979);
Nucl. Phys. {\bf B223}, 422, 433 (1983).}

\bibitem{other}{In principle, one could attempt an analogous generalization in
the electroweak sector, considering one or both of the factor groups
in this sector as members of an infinite sequence of groups, like
the generalization $SU(3) \to SU(N_c)$ in the QCD sector.  However, first,
there is less motivation for this, since a primary reason for the study of the
large-$N_c$ limit in QCD was to have an analytic method for nonperturbative
calculations, but perturbation theory is adequate to calculate  the observed
electroweak decays and reactions.  Second, such a generalization of the
electroweak sector is not straightforward; for example, if one extends the
weak isospin gauge group from SU(2) to $SU(N_{wk})$, $N_{wk} \ge 3$, one
encounters difficulties with anomalies.}

\bibitem{abj}{S. L. Adler, Phys. Rev. {\bf 177}, 2426 (1969); J. S. Bell and
R. Jackiw, Nuovo Cim. {\bf A60}, 47 (1969).}

\bibitem{gj}{D. J. Gross and R. Jackiw, Phys. Rev. {\bf D6}, 477 (1972).}

\bibitem{witten}{E. Witten, Phys. Lett. {\bf B117}, 324 (1982).}

\bibitem{gaugegrav}{R. Delbourgo and A. Salam, Phys. Lett. {\bf 40B}, 381
(1972); T. Eguchi and P. Freund, Phys. Rev. Lett. {\bf 37}, 1251 (1976);
L. Alvarez-Gaum\'e and E. Witten, Nucl. Phys. {\bf B234}, 269 (1983).}

\bibitem{bim}{C. Bouchiat, J. Iliopoulos, and P. Meyer, Phys. Lett. {\bf B38},
519 (1972).}

\bibitem{marsh}{C. Q. Geng and R. E. Marshak, Phys. Rev. {\bf D39}, 693
(1989); Phys. Rev. {\bf D41}, 717 (1990); K. S. Babu and R. N. Mohapatra,
Phys. Rev. {\bf D41}, 271 (1990);
J. A. Minahan, P. Ramond, and R. C. Warner, Phys. Rev. {\bf D41}, 716 (1990);
R. Foot, G. C. Joshi, H. Lew, and R. R. Volkas, Mod. Phys. Lett. {\bf A5},
2721.}

\bibitem{cy}{C.-K. Chow and T.-M. Yan, Cornell preprint 95/1378
(hep-ph/9512243).}

\bibitem{dim5}{Such a dimension-5 (gauge-invariant) operator is
\beq
{\cal O} = \frac{1}{M_X}\sum_{i,j}h_{ij}
(\epsilon_{ak}\epsilon_{bm}+\epsilon_{am}\epsilon_{bk})
\Bigl [ {\cal L}^{Ta}_{i L}C {\cal L}^b_{j L} \Bigr ] \phi^k \phi^m + h.c.
\label{majleft}
\eeq
where $i,j$ are generation indices, $a,b,k,m$ are SU(2) indices, and $\phi$
denotes the standard-model Higgs field or the $H_d$ Higgs field in a
supersymmetric extension of the standard model.
The term arising from the vev's of the
Higgs is a (left-handed, Majorana) neutrino mass term, which is naturally small
if $M_X$ is much larger than the electroweak symmetry-breaking scale.}

\bibitem{seesaw}{M. Gell-Mann, R. Slansky, and P. Ramond, in {\it
Supergravity}, (North Holland, 1979), p. 315; T. Yanagida, in
{\it Proceedings of the Workshop on Unified Theory and Baryon Number in the
Universe} (KEK, Japan, 1979).}

\bibitem{nutc}{The conventional seesaw mechanism and the dimension-5
operator in \cite{dim5} both presume that the theory has the requisite
fundamental Higgs fields.  If, instead, one assumes that electroweak
symmetry breaking occurs dynamically,
without fundamental Higgs fields, then Dirac neutrino
masses, like the masses of other fermions, would arise from multifermion
operators; however, if neutral $\nu_{iR}$ or $e_{iR}$ fields were present, as
in the respective cases $C5_\ell$ and $C4_\ell$ (eqs. (\ref{qnuzero}),
(\ref{qezero})), they would, of course,
still produce right-handed Majorana bare mass terms.}

\bibitem{add}{One could consider the $N_c$--extended standard model with
right-handed neutrinos and analyze the constraints on the quark hypercharges
while fixing the hypercharges of the leptonic fields to be equal to their
conventional values.  From the viewpoint of anomalies alone, this would be
unmotivated,
since all of the fermion hypercharges enter, {\it a priori}, on an equal
footing in the anomaly cancellation conditions.  Of course, if one
imposes the additional requirement beyond anomaly
cancellation that there exist (gauge-invariant) right-handed Majorana mass
terms, this implies that the leptonic hypercharges are automatically
fixed to their conventional values.
We shall not at the outset impose such a requirement here, since our purpose
is to explore the consequences of anomaly cancellation by itself.}

\bibitem{lep}{As is discussed later in the text, there are two exceptions to
this generic $Y_{\nu_R} \ne 0$ situation:
(i) case $C4_\ell$ in eq. (\ref{qezero})
in which the electron charge $q_e=0$, so that the electron-type leptons are
naturally light; and
(ii) case $C2_{q,sym}=C2_{\ell,sym}$ in which $q_\nu=-q_e=1/2$
(see text for details).}

\bibitem{n4}{
For completeness, on a purely empirical level, one must mention the
possibility of further generations with electroweak nonsinglet
neutrinos with masses $m_\nu > m_Z/2$ which would not be counted in the
LEP-SLC determination of $N_\nu$.}

\bibitem{alt}{
Here, we use the conventional approach of describing a baryon as a
color-singlet bound state of $N_c$ (valence) quarks; another useful
description, which is motivated by the large-$N_c$ limit, is to describe a
baryon as a soliton \cite{wittennc}.}

\bibitem{npert}{Recall that the nonperturbative violation of $B$ (and $L$)
in the
standard model, via electroweak instantons (G. 't Hooft, Phys. Rev. Lett.
{\bf 37}, 8 (1976)) is negligibly small at zero temperature.}

\bibitem{su2lgt}{I-H. Lee and R. Shrock, Phys. Rev. Lett. {\bf 59}, 14
(1987); Phys. Lett {\bf 196B}, 82 (1987); Nucl. Phys. {\bf B305}, 305 (1988);
R. Shrock,  Nucl. Phys. {\bf B4} (Proc. Suppl.), 373 (1988); S. Aoki, I-H. Lee,
and R. Shrock, Phys. Lett. {\bf 207B}, 471 (1988); {\it ibid.} {\bf 219B},
335 (1989).}

\bibitem{str}{It may be recalled that string theory provides an alternate
way of achieving gauge coupling unification (either without or with a simple
gauge group) via the relation $g_a^2 k_a = const.$, $a=1,2,3$ for the $U(1)_Y$,
$SU(2)$, and $SU(3)$ factor groups, where $k_a$ denotes the level of the
Kac-Moody worldsheet algebra \cite{gk}.  One has $k_2=k_3=1$ naturally (e.g. to
explain the absence of exotic representations of fermions); one can also argue
for $k_1=5/3$. However, in string theory the rank of
the gauge group is bounded above as $rank(G) \le 22$ \cite{ff}, thereby
precluding the discussion of $G_{SM}'$ with arbitrary $N_c$, in a string
context.}

\bibitem{gk}{P. Ginsparg, Phys. Lett. {\bf B197}, 139 (1987); V. Kaplunovsky,
Nucl. Phys. {\bf B307}, 145 (1988), errata, hep-th/9205070.}

\bibitem{ff}{H. Kawai, D. Lewellen, and S.-H. H. Tye, Phys. Rev. Lett.
{\bf 57}, 1832 (1986); Phys. Rev. {\bf D34}, 3794 (1986);
W. Lerche, D. L\"ust, and A. N. Schellekens, Nucl. Phys. {\bf B287}, 477
(1987);
I. Antoniadis, C. Bachas, and C. Kounnas, Nucl. Phys. {\bf B289}, 87 (1987).}

\bibitem{split}{Of course, in this discussion, we do not consider the
well-known problems faced by even supersymmetric GUT's, such as doublet-triplet
splitting in the Higgs chiral superfields.}

\bibitem{gg}{H. Georgi and S. L. Glashow, Phys. Rev. {\bf D6}, 429 (1972).}

\bibitem{hu}{S. T. Hu, {\it Homotopy Theory} (Academic Press, 1959).}

\bibitem{rem}{C. Q. Geng, R. E. Marshak, Z. Y. Zhao, and S. Okubo,
Phys. Rev. {\bf D36}, 1953 (1987); S. Okubo, C. G. Geng, R. E. Marshak, and
Z. Y. Zhao, {\it ibid.} {\bf D36}, 3268 (1987); S. Elitzur and V. P. Nair,
Nucl. Phys. {\bf B243}, 205 (1984).}

\bibitem{gggut}{H. Georgi and S. L. Glashow, Phys. Rev. Lett. {\bf 32}, 438
(1974).}

\bibitem{exc}{For the the special cases $C4_\ell$ and $C5_\ell$ which allow
removal of the $e_{iR}$ or $\nu_{iR}$ fields, respectively, the low-energy
theory has $N_f=4N_c+3$ Weyl matter fermion fields for each generation.
Thus $N_f$ is odd for both the cases $N_c$ odd and $N_c$ even.  Since the
spinor representations of the respective groups $SO(2N_c+4)$ and $SO(2N_c+2)$
for these cases are both even (both $=2^{N_c+1}$), it is not possible to
embed these fermion fields, by themselves, precisely in this
spinor representation.  The absence of a GUT for the case without $\nu_{iR}$'s
has been noted in Ref. \cite{cy}.  For both case $C4_\ell$ and $C5_\ell$ one
knows that it is necessary to add back the $G_{SM}'$-singlet
fields $e_{iR}$ or $\nu_{iR}$, respectively, to be able even to consider
fitting the fermions into the respective spinor representation.
One is thus led back to
the case where $N_f$ is given by eq. (\ref{nf}).}

\end{thebibliography}
\end{document}